\documentclass[aps,twocolumn,superscriptaddress,altaffilletter,lengthcheck,tightenlines]{revtex4}

\begin{document}
\title{INHOMOGENEOUS EXACT SOLUTIONS WITH
VARYING COSMOLOGICAL TERM}
\author{LUIS P. CHIMENTO}
\affiliation{Departamento de F\'{\i}sica, Universidad de Buenos Aires, \\
1428 Buenos Aires, Argentina\\
E-mail: chimento@df.uba.ar}
\author{DIEGO PAV\'{O}N}
\affiliation{Departamento de F\'{\i}sica, Universidad Aut\'onoma de Barcelona,\\
E-08193 Bellaterra (Barcelona), Spain\\
E-mail: diego@ulises.uab.es}

\begin{abstract}
We study the evolution of LTB Universe models possessing 
a varying cosmological term and a material fluid.
\end{abstract}

\maketitle

\section{Introduction}
There is at present an increasing feeling in the astrophysic
community that the cosmological constant
is not zero but should contribute substantially to the mass-energy 
of the Universe -see Weinberg {\it et al.} \cite{DHW} and references
therein. This may be so if the energy of the quantum vacuum spontaneously
decayed into matter and radiation, hence reducing the cosmological
term to a value compatible with astronomical constraints
-see for instance Overduin {\it et al.} \cite{JMO} and references 
therein. On the other hand, recently it has been pointed out that
because of sources evolution it may well happen that the Universe 
is in reality inhomogeneous and describable by the Lama\^{\i}tre-
Tolmann-Bondi (LTB) metric \cite{MHE}. Further motivations conductive
to use inhomogeneous metrics can be found in Krasinski \cite{KRS}.

\section{Metric and models} 
We consider a spatially flat LTB metric 
\begin{equation}
ds^{2} = - dt^{2} + Y^{'2} \, dr^{2} + Y^{2}\, (d\theta^{2} + sin^{2}
\theta \, d\phi^{2}), \; \; (Y = Y(r, t)) 
\label{1}
\end{equation}
whose source is a perfect fluid,
with equation of state $P = (\gamma - 1) \rho$, plus a varying
cosmological term $\Lambda(t)$.
The non-trivial Einstein equations are
\begin{equation}
\rho + \Lambda = \frac{1}{Y^{2} \, Y^{'}} 
(\dot{Y}^{2} \, Y)^{'} \, ,
\label{2}
\end{equation} 
\begin{equation}
P - \Lambda = - \frac{1}{Y^{2} \dot{Y}} (\dot{Y}^{2}\, Y)^{.} \, ,
\label{3}
\end{equation}
\begin{equation}
\frac{\ddot{Y}}{Y} + \left( \frac{\dot{Y}}{Y}\right)^{2} -
\frac{\ddot{Y}^{'}}{Y^{'}}-\frac{\dot{Y}}{Y}\frac{\dot{Y}^{'}}{Y^{'}}=0 
\qquad (8 \pi G = 1).
\label{4}
\end{equation}
In general the solutions can be expressed as
$Y(r, t) = R(r)^{2/3} Z(t)^{2/3\gamma} $. Next we summarize different
secenarios of interest -see Chimento and Pav\'{on} \cite{LD} for details.

\begin{enumerate}
\item For $\gamma$ and $\Lambda$ constants one obtains
\begin{equation}
Y_{1} = R^{2/3} (r) \, C_{1}^{2/3\gamma} \, cosh^{2/3\gamma} \left(
\frac{\sqrt{3 \gamma \Lambda}}{2} \; t + \varphi_{1} \right) \, , \\
\label{5}
\end{equation} 
\begin{equation}
Y_{2} = R^{2/3} (r) \, C_{2}^{2/3\gamma} \, sinh^{2/3\gamma} \left(
\frac{\sqrt{3 \gamma \Lambda}}{2} \; t + \varphi_{2} \right).
\label{6}
\end{equation}
Obviously both sets of solutions have a final inflationary stage.
\item
When $\gamma = \mbox{constant}$ and 
\begin{equation}
\Lambda (t) = \frac{\lambda_{0}^{2}}{t^{2}} \quad
(\lambda_{0}^{2} = \mbox{constant}) \, ,
\label{7}
\end{equation}
it follows that
\begin{equation} 
Z(t) =C_{1} \, t^{m_{+}} + C_{2} \, t^{m_{-}} 
\label{8}
\end{equation} 
\  \\
where $ m_{\pm} =  \left(1 \, \pm \, \sqrt{1 + 3 \gamma
\lambda_{0}^{2}}\right)/2 $. Inflationary solutions 
may occur for large enough $\lambda_{0}^{2}$.
\item
For $\gamma = \mbox{constant}$ and 
\begin{equation}
\Lambda = \lambda_{0}^{2} \, t^{n-2}  \; \; \; (n \neq 0, \, 2) \, ,
\label{9}
\end{equation}
the solution can be expressed as a combination of Bessel functions
$$
Z = C_{1} \, t^{1/2} \, J_{1/n}\left(\frac{\lambda_{0}}{n}
\sqrt{- 3\gamma} \, t^{n/2} \right)
$$
\begin{equation}
+ C_{2}  \, t^{1/2} \, J_{-1/n}\left(\frac{\lambda_{0}}{n}
\sqrt{- 3\gamma} \, t^{n/2} \right).
\label{10}
\end{equation}
The behavior at the asymptotic limits depends on $n$. 
For $0 < n <2$ one has the following: (i) When 
$ t \rightarrow  0$ one obtains $Z \sim C_{1} \, t + C_{2}$
-one can choose $C_{2} = 0$ to
have the initial singularity at $t = 0$. (ii) When 
$t \rightarrow \infty$ there follows
$Z \sim t^{\frac{1}{2}-\frac{n}{4}} \; cos \, t^{n/2}$.\\
\noindent Likewise for $ n < 0 $: 
(i) when $ t \rightarrow 0 \; $ one obtains
$Z \sim t^{\frac{1}{2}-\frac{n}{4}} \; cos \, (t^{n/2} + \varphi)\;. $
(ii) When $ t \to \infty \; $ one obtains
$ Z \sim t \, .$

\item For $\gamma = \mbox{constant}$ and
\begin{equation}
\Lambda (t) =\lambda_{0}^{2}+c\mbox{e}^{-\alpha t} \qquad (c < 0)\, ,
\label{11}
\end{equation}
where $\lambda_{0}^{2}$, $\alpha$ and $c$ are constants,
again the general solution is a combination of Bessel functions
$$
Z = C_{1} \,J_{\frac{\lambda_{0}}{\alpha}\sqrt{3\gamma}}
\left(\frac{\sqrt{- 3\gamma c}}{\alpha}\,\mbox{e}^{\frac{-\alpha}{2} t}\right)
$$
\begin{equation} 
+ \, C_{2} \,J_{-\frac{\lambda_{0}}{\alpha}\sqrt{3\gamma}}
\left(\frac{\sqrt{- 3\gamma c}}
{\alpha}\,\mbox{e}^{\frac{-\alpha}{2} t}\right) \, ,
\label{12}
\end{equation} 
with 
\begin{equation}
C_2=-\frac{J_{\frac{\lambda_{0}}{\alpha}\sqrt{- 3\gamma}}
\left(\frac{\sqrt{- 3\gamma c}}{\alpha}\right)}
{J_{-\frac{\lambda_{0}}{\alpha}\sqrt{- 3\gamma}}
\left(\frac{\sqrt{- 3\gamma c}}{\alpha}\right)}\,C_1
\label{13}
\end{equation}
in order to fix the initial singularity at $t = 0$.
When $t \rightarrow 0$ one has $Z \sim t$. 
At the final stage, when $t \rightarrow \infty$ 
and $\Lambda \rightarrow \lambda_{0}^{2}$, one
obtains the following asymptotic behavior
\begin{equation}
Y\approx R^{2/3}(r)\,\mbox{e}^{\frac{\lambda_{0}}
{\sqrt{-3\gamma c}}\,t} \, .
\label{14}
\end{equation}

For the particular case $\lambda_{0}^{2} = 0 $ and
in the limit $t \rightarrow \infty$, there is a solution
whose  final behavior is
\begin{equation}
Y\approx R^{2/3}(r)\,t^{2/3\gamma} \, .
\label{15}
\end{equation}

\item For $\gamma = \gamma (t)$ and $\Lambda = \Lambda (t)$
it can be found expressions for both quantities, 
\begin{equation}
\Lambda(t)=\frac{4C^2(t-t_0)^{2n}}{3\gamma_0 n^2(n+1)^2}
\left[1+\frac{(t-t_0)^{n+1}}{C}\right]^{\frac{2-n}{n}},
\label{16}
\end{equation}
\begin{equation}
\gamma(t)=\gamma_0\left[1+\frac{(t-t_0)^{n+1}}{C}\right]^{-\frac{2+n}{n}},
\label{17}
\end{equation}
as well as an asymptotic solution for $Y(t,r)$
\begin{equation}
Y\approx R^{2/3} (r) \, T_{0}^{2/3\gamma_0} \,
\left[\frac{(n+1)(n+2)}{n}(t-t_0)\right]^{2/3\gamma_0} \, ,
\label{18}
\end{equation}
where $T_{0}, \gamma_{0}, t_{0}, \mbox{and} C $ are constants.
It is worthy of note that, for $t\gg t_0$ we
have both $\gamma \rightarrow \gamma_0$ and 
$\Lambda \rightarrow 0$. 
\end{enumerate}

To examine the singular structure of the plane LTB metric (1) 
we have calcuated the curvature scalar 
and evaluated it at the points where the coefficients of 
the metric $Y^{'2}$ and/or $Y^{2}$ vanish. All the solutions 
we have found for $\gamma = \mbox{constant}$ except (\ref{5}) have a 
singularity at $t = 0$, i.e. the big-bang singularity.

\section{Conclusions}
We have found the coefficients of the 
LTB  metric assuming that the early
Universe possessed a time varying cosmological term, and
that the adiabatic index of the material fluid were either 
constant or not. \\
\noindent (a) All the solutions we have derived contain
an arbitrary function of the radial coordinate. \\
\noindent (b) For 
$\gamma = $ constant all the solutions, except (\ref{5})
have a singularity at $t = 0$, i.e. the big-bang singularity.\\
\noindent (c) Constant as well as varying cosmological terms give
rise asymptotically to exponential inflation -see (\ref{5}),
(\ref{6}) and (\ref{14}).\\
\noindent (d) For $\Lambda(t) \propto e^{-\alpha t}$
there exist solutions which behave as though 
the Universe were asymptotically matter dominated 
at late times when $\gamma = 1$, i.e. $Y \propto t^{2/3}$. 

\section*{Acknowledgements}
This work was partially supported by the Spanish Ministry of
Education under Grant PB94-0718, and the University of Buenos Aires
under Grant EX-260.

\end{document}